# Cantilever-Based Biosensors in CMOS Technology

K.-U. Kirstein, Y. Li, M. Zimmermann, C. Vancura, T. Volden, W. H. Song, J. Lichtenberg, and A. Hierlemannn,
Physical Electronics Laboratory, ETH Zurich, Switzerland

## Abstract

*Single-chip CMOS-based biosensors that feature microcantilevers as transducer elements are presented. The cantilevers are functionalized for the capturing of specific analytes, e.g., proteins or DNA. The binding of the analyte changes the mechanical properties of the cantilevers such as surface stress and resonant frequency, which can be detected by an integrated Wheatstone bridge. The monolithic integrated readout allows for a high signal-to-noise ratio, lowers the sensitivity to external interference and enables autonomous device operation.*

## 1. Introduction

Many methods for detecting biological substances in a solution are based on the use of fluorescent markers and the corresponding optical analysis. This method is very time consuming and needs a complex and expensive optical setup. For the increasing number of biochemical analyze procedures in the daily healthcare, like blood analysis for antibodies or other proteins, a fast, easy-to-use and cheaper alternative to fluorescent methods is desired.

CMOS-based microsystems can provide such easy-to-handle biochemical tests. In this paper, two microcantilever-based approaches are presented, which build the base for a variety of different applications in medical diagnoses. Specific analyte detection is achieved by taking advantage of bio-affinity recognition between the analyte and a suitable probe molecule, e.g. immunoassay. For the detection of a specific antigen in the patient's sample, the corresponding antibody is immobilized on the cantilever surface prior to the actual analysis. Once in contact with the sample the analyte is specifically captured, which results in a change of cantilever properties [1].

The first approach uses a static cantilever, sensing the static bending of the cantilever, which is due to an increased surface stress caused by the adhering analyte (see Figure 1). Another approach uses cantilevers that are mechanically excited at their resonant frequency. The additional mass of the analyte molecules causes a shift in the resonant frequency upon binding (Figure 2). In both cases, the mechanical deformation of the cantilever is sensed by an integrated piezoresistive Wheatstone bridge. Monolithic integrated readout circuitry allows for on-chip amplification of the weak bridge voltage, leading to a high sensitivity of the device.

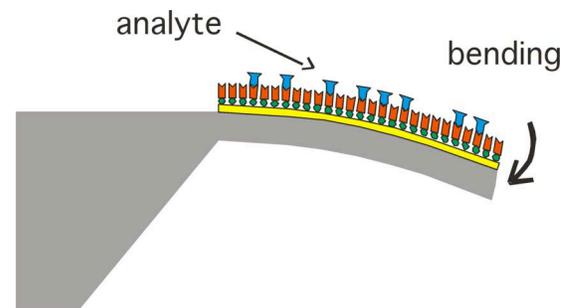

**Figure 1: Bending of a static microcantilever due to analyte binding**

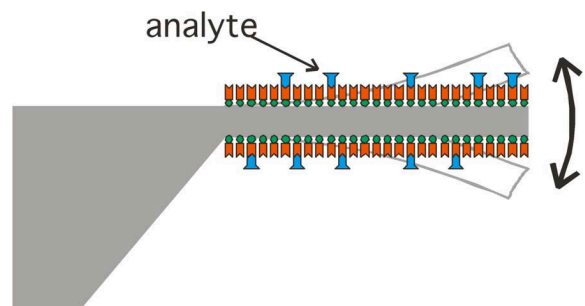

**Figure 2: Resonant operation of the microcantilever**

## 2. Fabrication

The cantilever-based biosensors are fabricated in a standard 0.8μm double-poly, double-metal CMOS process with post-CMOS micromachining. After completion of the CMOS process, a back-side anisotropic silicon etch is performed using potassium hydroxid (KOH) together with an electro-chemical etch-stop. The pn-junction for this etch-stop is defined by the n-well diffusion layer of the CMOS-technology, providing a well-defined thickness of the crystalline silicon layer forming the cantilever [2]. The cantilever is released by two successive anisotropic front-side dry etch steps, which remove the dielectric layers and the bulk silicon, respectively. Figure 3 shows the cantilever structure before and after the post-processing. The design of the three additional mask layers is completely integrated in the physical design flow of the CMOS technology, so that the physical design verification, e.g., design-rule checks, can be performed with respect to the CMOS layers. The complete post-processing can be performed on wafer level, leading to a very cost-efficient mass-production.

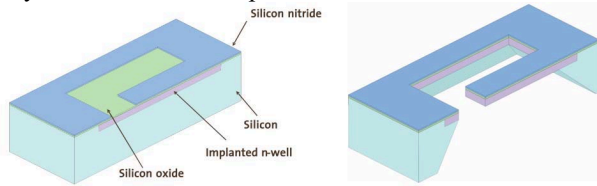

a) before post-processing    b) after post-processing

**Figure 3: Schematic view of the cantilever structure, before and after post-processing**

## 3. System Architecture

The piezoresistive Wheatstone bridge for the readout of the resonant oscillation is placed on the clamped edge of the cantilever, where the maximum mechanical stress is induced. In case of the static system this measurement bridge is distributed of the cantilever length and covers a larger area.

### 3.1 Static Cantilever Readout Circuit

Figure 4 shows a block diagram of the readout circuitry for the static cantilever system. An array of four cantilevers is connected to the readout amplifiers by an analog multiplexer. A chopper-stabilized amplifier as first stage performs a low-noise, low-offset amplification of the weak sensor signal. This first stage is followed by a low-pass filter to improve the signal-to-noise ratio, a programmable offset compensation stage and two additional gain stages.

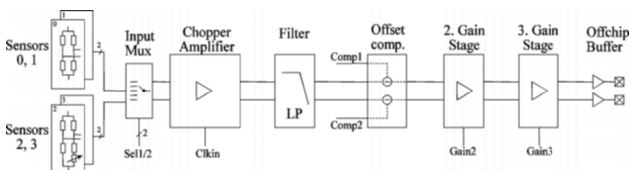

**Figure 4: Block diagram of the readout circuit for static cantilever operation**

### 3.2 Resonant Cantilever Feedback Circuit

The readout and feedback circuit for resonant operation of the microcantilevers is shown in Figure 5. The actuation of the cantilever is performed by a coil along the cantilever edges, driven by a periodic electric current with a frequency equal to the mechanical resonance of the cantilever. Together with a permanent magnet, integrated in the package of the sensor chip, the acting Lorentz force leads to a bending of the cantilever [3].

In case of a resonant cantilever operation, the piezoresistive Wheatstone bridge has been accomplished by p-channel MOS transistors biased in the linear region, which has the advantage of a higher resistivity and lower power consumption compared to diffusion-type silicon resistors. As can be seen from Figure 5, a feedback loop has been designed in order to stabilize the resonant mode of the cantilever. The first amplifier stage is a low-noise, fully differential instrumentation amplifier using a fully differential-difference amplifier (DDA) in a non-inverting feedback configuration. High-pass filters in the feedback loop improve the signal-to-noise ratio by damping the low-frequency noise originating in the MOS-based Wheatstone bridge. A variable gain amplifier allows to adjust to different mechanical damping of the cantilever, due to different liquids presented to the biosensor. A non-linear amplifier limits the amplitude of the feedback loop for stable operation and drives the low-resistance coil via a class AB output buffer. The readout block mainly consists of a digital counter to monitor the resonant frequency of the sensor system.

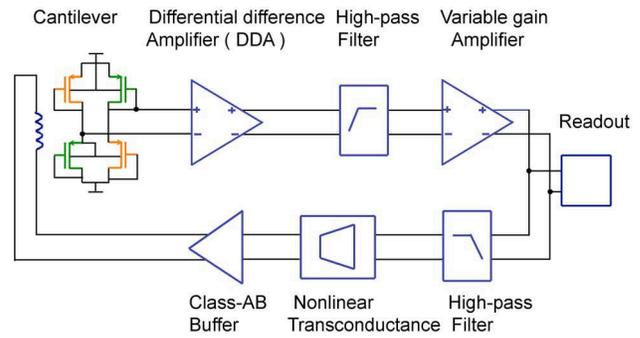

**Figure 5: Block diagram of the feedback circuitry for resonant cantilever systems.**

## 4. Acknowledgements

The authors thank Prof. Henry Baltes (on leave) for sharing laboratory resources and for his scientific advice.